\documentclass[journal]{vgtc}         

\ifpdf
  \pdfoutput=1\relax                   
  \usepackage{graphicx}                
  \DeclareGraphicsExtensions{.pdf,.png,.jpg,.jpeg} 
\else
  \usepackage{graphicx}                
  \DeclareGraphicsExtensions{.eps}     
\fi%

\graphicspath{{figures/}{pictures/}{images/}{./}} 

\usepackage{microtype}                 
\PassOptionsToPackage{warn}{textcomp}  
\usepackage{textcomp}                  
\usepackage{mathptmx}                  
\usepackage{times}                     
\usepackage{cite}                      
\usepackage{tabu}                      
\usepackage{booktabs}                  

\usepackage{enumitem}
\setlist[itemize]{noitemsep, topsep=0pt}
\usepackage{wrapfig}
\usepackage{caption}
\usepackage{titlecaps}
\usepackage{multirow}
\usepackage{dblfloatfix}
\usepackage{amsmath}
\usepackage{amssymb}
\usepackage{newtxmath}
\usepackage{physics}

\newcommand{\RNum}[1]{\uppercase\expandafter{\romannumeral #1\relax}}

\newcommand{\cmo}{\textcolor[rgb]{0, 0, 0}}
\newcommand{\para}[1]{\vspace{1mm}\noindent\textbf{#1}}

\newcommand{\system}{\textsc{RL-Label}}
\newcommand{\IDE}{\texttt{Identify}}
\newcommand{\COM}{\texttt{Compare}}
\newcommand{\SUM}{\texttt{Summarize}}
\newcommand{\STU}{STU}
\newcommand{\NBA}{NBA}
\newcommand{\Ours}{\textsc{Ours}}
\newcommand{\Force}{\textsc{Force}}
\newcommand{\No}{\textsc{No}}

\onlineid{1532}

\vgtccategory{Research}
\vgtcpapertype{algorithm/technique}

\title{\system{}: A Deep Reinforcement Learning Approach \cmo{Intended} for\\ AR Label Placement in Dynamic Scenarios}

\author{Chen Zhu-Tian$^1$, Daniele Chiappalupi$^{1, 2}$, Tica Lin$^1$, Yalong Yang$^3$, Johanna Beyer$^1$, Hanspeter Pfister$^1$}
\authorfooter{
\item $^1$ Harvard John A. Paulson School of Engineering and Applied Sciences
\item $^2$ ETH Zurich, Zurich
\item $^3$ Virginia Tech
}

\shortauthortitle{Biv \MakeLowercase{\textit{et al.}}: Global Illumination for Fun and Profit}

\abstract{Labels are widely used in augmented reality (AR) to display digital information. Ensuring the readability of AR labels requires placing them in an occlusion-free manner while keeping visual links legible, especially when multiple labels exist in the scene. Although existing optimization-based methods, such as force-based methods, are effective in managing AR labels in static scenarios, they often struggle in dynamic scenarios with constantly moving objects. This is due to their focus on generating layouts optimal for the current moment, neglecting future moments and leading to sub-optimal or unstable layouts over time. 
In this work, we present \system{}, a deep reinforcement learning-based method \cmo{intended for managing the placement of AR labels} in scenarios involving moving objects. 
\system{} considers both the current and predicted future states of objects and labels, such as positions and velocities, as well as the user’s viewpoint, to make informed decisions about label placement. It balances the trade-offs between immediate and long-term objectives. 
\cmo{We tested \system{} in simulated AR scenarios} on two real-world datasets, 
showing that it effectively learns the decision-making process for long-term optimization, outperforming two baselines (i.e., no view management and a force-based method) by minimizing label occlusions, line intersections, and label movement distance. 
Additionally, a user study involving 18 participants indicates that, \cmo{within our simulated environment}, \system{} excels over the baselines in aiding users to identify, compare, and summarize data on labels in dynamic scenes.
} 


\CCScatlist{ 
 \CCScat{K.6.1}{Management of Computing and Information Systems}%
{Project and People Management}{Life Cycle};
 \CCScat{K.7.m}{The Computing Profession}{Miscellaneous}{Ethics}
}

\teaser{
  \centering
  \hspace*{-1.8cm}
  \includegraphics[width=1.2\linewidth]{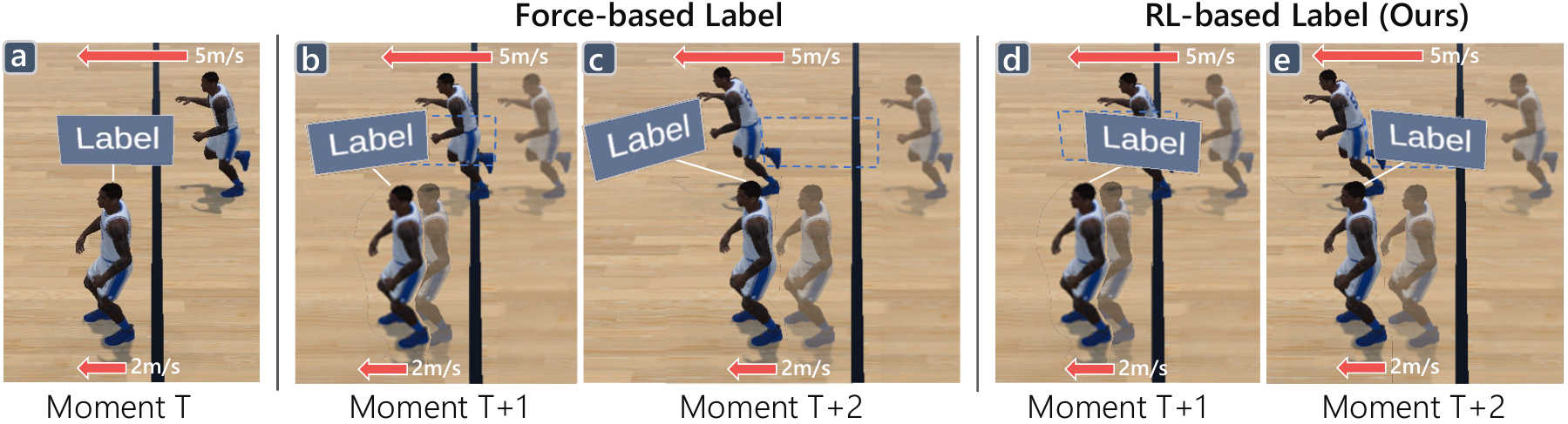}
  \caption{
    \system{} not only adapts label placement considering players' current motion status (e.g., speed, direction) and long-term outcomes but also ensures label stability over time.
    (a) Both players move left, with the rear player moving faster. A label is attached to the front player.
    (b)-(c) Using a force-based method, the label shifts left to avoid immediate occlusion but results in future occlusion.
    (d)-(e) With our method, the label moves right, sacrificing some immediate occlusion-free space for preventing future occlusion and ensuring stable visibility.
  }
  \label{fig:teaser}
}

\vgtcinsertpkg

\begin{document}


\firstsection{Introduction}

\firstsection{Introduction}

\maketitle


Many augmented reality (AR) applications (e.g., sports analytics~\cite{lin2021towards, DBLP:conf/chi/ChenYSLBXP23}, 
mechanical maintenance~\cite{DBLP:conf/chi/MohrMLTSK20}, 
and education~\cite{DBLP:journals/tvcg/TongCXLYBQ23, DBLP:conf/chi/ChenTWBQ20})
use labels to display digital information.
A label is a small virtual canvas connected to physical objects via leader lines.
Determining label layouts is essential to help users perceive the visual information.
Thus, extensive research has explored automatic generation of AR label layouts, with most methods treating it as an optimization problem~\cite{DBLP:journals/cgf/BekosNN19}.
These view management systems aim to maximize objectives
such as preventing label-object overlaps, 
avoiding leader line intersections, 
and minimizing distances between labels and their physical referents~\cite{DBLP:journals/cgf/BekosNN19}.
Though these methods work well for static scenarios, 
they often fall short in dynamic scenarios where physical objects, like athletes in sports games, are constantly moving. 
This presents a clear need for improved AR label management systems that can adapt to changing object positions.



Managing label layouts for dynamic objects in AR presents unique challenges. Labels need to adapt their positions continually and stably to moving objects while avoiding occlusion and line intersections. 
Thus, view management systems should generate layouts that are optimal not only for the current moment but also capable of transitioning smoothly to future layouts. 
However, existing methods primarily focus on optimizing layouts for the present, often neglecting future scenarios. 
This leads to sub-optimal or unstable layouts over time.
For instance, 
one most widely-used method for label management is the force-based algorithm~\cite{DBLP:journals/cgf/BekosNN19}, 
which models objectives as forces (e.g., occlusion-free placement is modeled as repulsive forces between labels and objects). 
Consider a scenario where both players in~\autoref{fig:teaser}a are moving to the left, 
with the rear player moving faster and a label attached to the front player. 
A force-based method would push the label to the left to avoid occluding the rear player (\autoref{fig:teaser}b).
However, this would result in severe future occlusion since the rear player also moves to the left (\autoref{fig:teaser}c).

We approach the AR labels placement problem for moving objects from a new angle:
the view management system should generate label layouts based on both current and predicted future states.
Inspired by other applications in dynamic real-world environments (e.g., self-driving cars, robotics control),
we achieve this goal by employing a reinforcement learning (RL) method.
Specifically, we propose \system{}, 
an RL-based view management system for scenes with multiple moving objects, each with an attached label. 
To the best of our knowledge, we are the first to explore using RL to manage AR labels for moving objects.
\system{} observes object and label states, such as positions and velocities, and the user's viewpoint, 
to make informed decisions about label placement
that balance the trade-offs between immediate and long-term objectives.
For instance, \system{}  moves the label to the right, 
while temporarily causing occlusion of the players (\autoref{fig:teaser}d), 
to prevent future occlusion (\autoref{fig:teaser}e).

\cmo{To ensure reproducibility, we tested \system{} within simulated AR environments based on Virtual Reality and two real-world trajectory datasets (one for NBA players and the other for students on a campus).}
Computational experiments demonstrate that \system{} can effectively learn the decision process to achieve a long-run optimization,
outperforming two baselines (i.e., no view management and a force-based method) by reducing occlusions, line intersections, and movement distance of the labels.
Furthermore, we conducted a formal user study with 18 participants 
\cmo{in the same simulated environments}
to compare \system{} with the two baselines
in assisting users in three visual search tasks, i.e., identify, compare, and summarize data on AR labels.
Overall, 
\cmo{within the simulated environments,}
users performed these tasks faster, more accurately, and with a lower mental load when using \system{} than the baselines.

In summary, our contributions are threefold:
First, we formulate the label placement in dynamic scenes as a sequential decision (instead of an optimization) problem with the goal to maximize cumulative rewards;
Second, we design and develop \system{}, an RL-based method to solve this problem;
Third, we evaluate \system{} with both computational experiments and user studies,
showing that it outperforms force-based methods in both quantitative and qualitative aspects.
\section{Related Work}


\vspace{-1mm}
\para{Labels in Augmented Reality.}
Labels have long been used to annotate objects in illustrations~\cite{DBLP:journals/cgf/BekosNN19}.
Since the early '80s,
automated view management systems for label placement have emerged in computer graphics research~\cite{ahn1984part},
varying by label type (\emph{internal} or \emph{external}) and environment (\emph{desktop} or \emph{AR}). 
This work concentrates on placing external labels in AR environments.

\begin{figure}[h]
  \centering
  \vspace{-7mm}
  \includegraphics[width=0.85\linewidth]{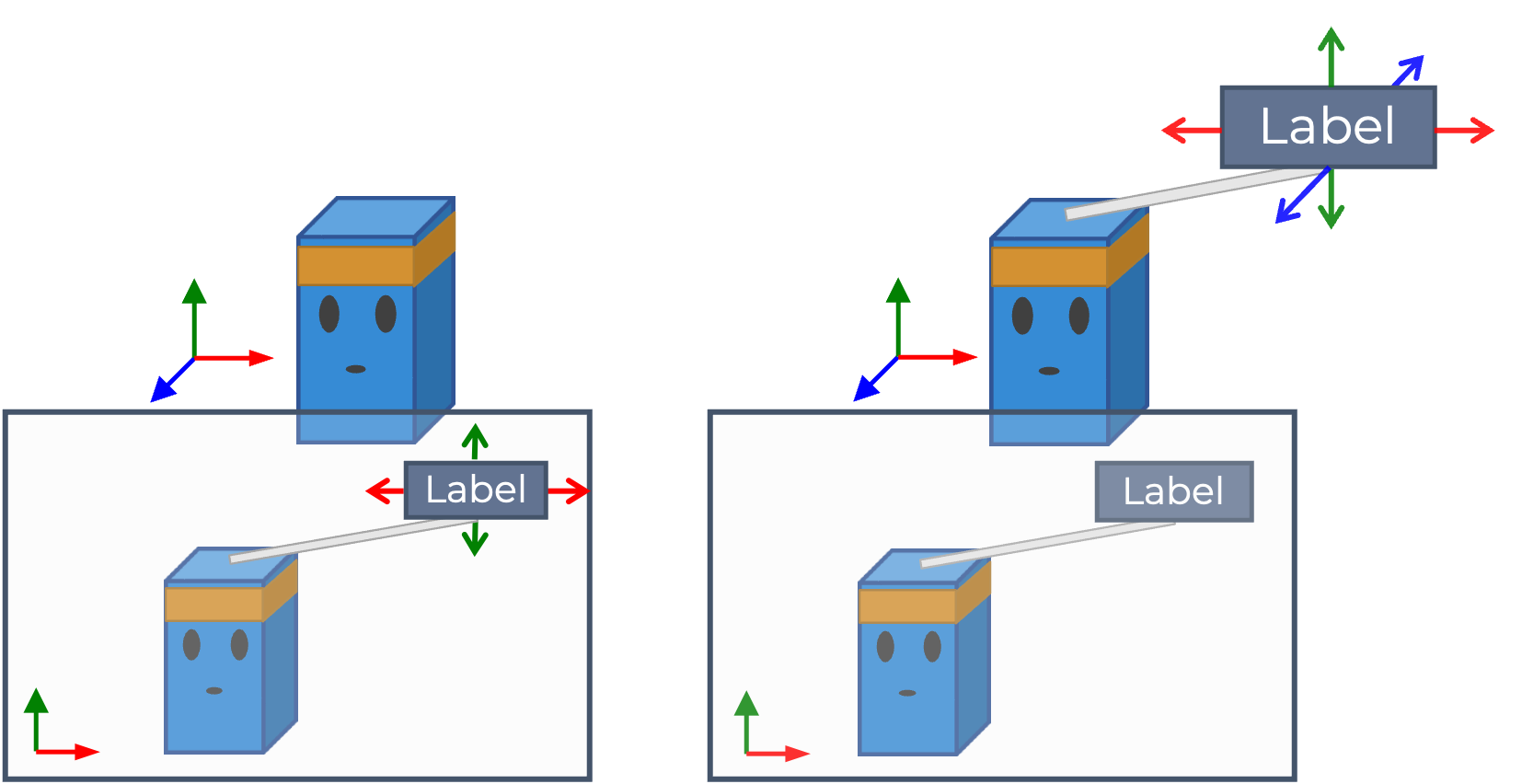}
  \caption{
    Left: 2D labels are placed and managed in the image space.
    Right: 3D labels are placed and managed in the world space, and subsequently projected onto the image plane.
  }
  \vspace{-3mm}
  \label{fig:2d_vs_3d}
\end{figure}

Early view management systems for AR environments mainly place labels in 2D screen space (\autoref{fig:2d_vs_3d} Left), unaffected by 3D transformations.
Representative examples include Azuma and Furmanski's identification of label clusters in screen space~\cite{DBLP:conf/ismar/AzumaF03}, 
\cmo{visual saliency driven label placement~\cite{DBLP:conf/ismar/GrassetLKTS12, DBLP:conf/icip/RakholiaHH18, DBLP:conf/wacv/HegdeMHK20}}, 
and Tatzgern et al.'s implementation of adaptive clustering based on information density to reduce visual clutter~\cite{DBLP:conf/vr/TatzgernOKJGS16}.
Yet, 2D labels face the ``floating labels'' issue~\cite{DBLP:conf/vr/TatzgernKGS14}, 
where frequently changing user viewpoints 
lead to unpredictable label positions due to changing projected 3D points during camera movements.

Unlike 2D labels, 3D labels (\autoref{fig:2d_vs_3d} Right) 
offer greater stability in AR when users change viewpoints. 
Tatzgern et al.~\cite{DBLP:conf/vr/TatzgernKGS14} propose Hedgehog labeling, a representative example of 3D labels. 
They use 3D geometric constraints to generate label placements that fulfill the desired objectives (e.g., occlusion-free) and are stable over time, even when the viewpoint changes.
Madsen et al.~\cite{DBLP:journals/tvcg/MadsenTMSK16} later compared this method to 2D labels in an empirical study.
Findings revealed that 3D labels with a limited update rate outperform 2D and continuously updated 3D labels.
More recently, Koppel et al.~\cite{DBLP:conf/apvis/KoppelGW21} contributed a system for 3D AR labels that manages label visibility and level of detail, 
considering both the labels in front of and out of the view. 
Gebhardt et al.~\cite{DBLP:conf/uist/GebhardtHOWHHB19} utilize RL to control the visibility (i.e., show or hide) of an object's label based on the user's gaze data.
Lin et al.~\cite{DBLP:journals/tvcg/LinYBP23} studied the design space of 2D and 3D AR labels for out-of-view objects.

While existing methods have proven useful and effective in various AR scenarios, they primarily focus on static objects with fixed positions.
However, in real-world environments,
objects often move dynamically (e.g., basketball players, vehicles, conveyor belt sushi).
This presents additional challenges for label placement,
which should consider both the current and predicted future states of the objects.
We aim to develop a view management system that observes dynamic moving objects and adapts 3D labels in real-time.

\para{Adaptive User Interfaces in Augmented Reality.}
A label is a type of user interface (UI) that displays visual content in a small canvas~\cite{DBLP:journals/cgf/BekosNN19}.
Unlike traditional desktop UIs, 
many design decisions of AR UIs, 
such as display location and manner, 
cannot be predetermined and must adapt to the user's context in real-time~\cite{DBLP:journals/tvcg/GrubertLZR17}.
Extensive research has focused on making AR UIs adaptive by leveraging geometry information from the environments.
For example, AR UIs can be aligned with edges~\cite{DBLP:conf/chi/NuernbergerOBW16},
placed on surfaces~\cite{DBLP:journals/tvcg/ChenS0WQW20}, 
and interact with 3D meshes~\cite{DBLP:conf/uist/FenderLHA017, DBLP:conf/chi/FenderHA018} extracted from the physical environments.

In addition to basic geometry information, 
recent AR UIs leverage semantic information of the scene to 
enhance the user experience.
For instance, Tahara et al.~\cite{DBLP:conf/ismar/TaharaSNI20} employ a scene graph to define the spatial relationships between virtual and physical objects, automatically adjusting the virtual content when the user moves to other environments.
AdapTutAR~\cite{DBLP:conf/chi/HuangQWPSCRQ21}, an AR task tutoring system, adjusts the teaching content adaptively based on the user's characteristics.
SemanticAdapt~\cite{DBLP:conf/uist/ChengY0SL21} uses computer vision techniques to detect the category of real objects and associates them with AR content.
Lindlbauer et al.~\cite{DBLP:conf/uist/LindlbauerFH19} control the placement and level of detail of virtual content by considering both the indoor environment and the cognitive load of the user's performing task.
ScalAR~\cite{DBLP:conf/chi/QianHHWIR22} enables designers to author semantically adaptive AR experiences in VR.
However, all these works focus on static scenarios where
the physical objects are placed in fixed positions.
In contrast, we 
target dynamic scenarios where objects are moving.

\para{AR Labeling as a Partially Observable Decision Problem.}
Fundamentally, managing labels for moving objects in AR involves dealing with a partially observable decision problem~\cite{DBLP:conf/nips/JaakkolaSJ94}.
In a partially observable environment, 
the entire state is partially visible to external sensors.
For instance, an object's destination and planned route are known only to the object itself.
This distinguishes AR from other 3D environments, where the system state is fully visible.
Recently, deep RL has demonstrated promising performance in addressing partially observable problems
in applications such as self-driving cars~\cite{DBLP:journals/tits/KiranSTMSYP22}, robotics control~\cite{DBLP:journals/robotics/LiuNZMD21}, 
and video games~\cite{DBLP:journals/corr/abs-1912-10944}.
Inspired by these applications, this work explores using deep RL to manage labels for moving objects.

\para{Reinforcement Learning for Data Visualizations.}
Unlike supervised learning, which relies on historical data, 
RL models learn from interactive experiences with the environment (e.g., play GO),
making RL suitable for dynamic scenarios 
requiring long-term optimization.

Only a few studies have explored RL methods for solving visualization problems. 
For instance, PlotThread~\cite{DBLP:journals/tvcg/TangLWLKKYREW21} uses RL to create user-preferred storyline layouts. 
MobileVisFixer~\cite{DBLP:journals/tvcg/WuTDLIQ21} employs RL methods to adapt visualizations for different mobile devices by adjusting parameters like size, offset, and margin. 
Table2Chart~\cite{DBLP:conf/kdd/ZhouLHLLJHCJZ21} uses deep Q-learning to recommend chart templates based on input data tables.
In summary, most existing works focus on static 2D environments.
We aim to apply RL-based methods to manage label layouts in dynamic 3D environments, an underexplored area with unique challenges.




\section{Problem Formulation}

This work, 
similar to Yao et al.~\cite{DBLP:journals/tvcg/YaoBVI22},
\cmo{focuses on scenarios where the viewer is stationary but can rotate their viewpoint horizontally or vertically} to observe multiple moving objects, such as watching sports in a stadium or monitoring environments using surveillance cameras.
We constrain that each object is annotated by a single label
and leave multiple-label or moving viewpoint scenarios for future research.

\begin{figure}[h]
  \centering
  \includegraphics[width=1\linewidth]{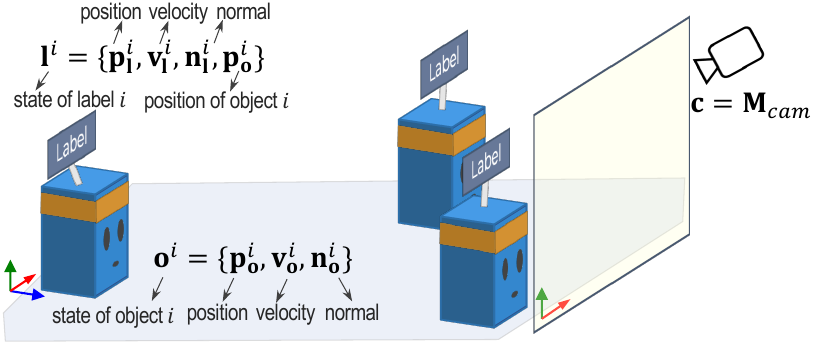}
\vspace{-2mm}
  \caption{We consider scenarios where a stationary viewer observes multiple moving objects, each with a corresponding label.}
  \label{fig:problem}
\end{figure}

\noindent
At each time step,
the view management process involves three components (\autoref{fig:problem}):
\begin{itemize}
    \vspace{1mm}
    \item \textbf{Object:} 
    An object $\mathbf{o}^i$ can be approximated as a cube shape rigid body 
    in the 3D \emph{world} space $\mathbb{R}^3_{wld}$.
    The object $\mathbf{o}^i$ attempts to move to 
    a goal position following a planned route and a preferred velocity.
    For a view management system,
    the observable state of the object     
    can be represented as 
    $\mathbf{o}^{i} = \{\mathbf{p}^{i}_\mathbf{o}, \mathbf{v}^{i}_\mathbf{o}, \mathbf{n}^{i}_\mathbf{o}\}$, 
    where $\mathbf{p}^{i}_\mathbf{o} = (x^{i}, y^{i}, z^{i})$,
    $\mathbf{v}^{i}_\mathbf{o} = (v_x^{i}, v_y^{i}, v_z^{i})$,
    and $\mathbf{n}^{i}_\mathbf{o} = (n_x^{i}, n_y^{i}, n_z^{i})$
    are its position, velocity, and normal, respectively.
    The observable state of all the objects 
    is thus $\mathbf{o} = \{\mathbf{o}^{i} \}^n_{i=1}$,
    where $n$ is the number of objects.
    
    \vspace{1mm}    
    \item \textbf{Label:} 
    A label $\mathbf{l}^i$ is a rectangle shape canvas in the 3D \emph{world} space $\mathbb{R}^3_{wld}$, 
    linking to its target object $\mathbf{o}^i$ through a leader line.
    The observable state of the label
    can be defined as
    $\mathbf{l}^{i} = \{\mathbf{p}^{i}_\mathbf{l}, \mathbf{v}^{i}_\mathbf{l}, \mathbf{n}^{i}_\mathbf{l}, \mathbf{p}^{i}_{\mathbf{o}} \}$,
    where $\mathbf{p}^{i}_\mathbf{l}$, $\mathbf{v}^{i}_\mathbf{l}$, and $\mathbf{n}^{i}_\mathbf{l}$
    are its position, velocity, and normal, respectively; $\mathbf{p}^{i}_{\mathbf{o}}$ represents the other endpoint of the leader line.
    In practice, 
    the label usually keeps its normal $\mathbf{n}^{i}_\mathbf{l}$
    pointing to the camera to maximize its readability.
    Similarly, 
    the state of all labels 
    is represented by $\mathbf{l} = \{\mathbf{l}^{i} \}^n_{i=1}$, 
    where $n$ is the number of labels, which equals the number of objects.

    \vspace{1mm}
    \item \textbf{Viewpoint:}
    A viewpoint is determined by the camera $\mathbf{c}$ and specifies the observer's position in relation to the objects and labels being watched. 
    Thus, the state of the viewpoint $\mathbf{c}$ 
    can be represented as the projection matrix $\mathbf{M}_{cam}$.

\end{itemize}

For a label $\mathbf{l}^{i}$,
a view management system should generate 
an action $\mathbf{a}^{i}$ to update its position
based on its current state $\mathbf{l}^{i}$,
the states of its target object $\mathbf{o}^{i}$,
other labels $\mathbf{l} \setminus \{\mathbf{l}^{i}\}$,
other objects $\mathbf{o} \setminus \{\mathbf{o}^{i}\}$,
and the viewpoint $\mathbf{c}$.
We simplify these states as
$\mathbf{s}^{i} = \{
    \mathbf{l}^{i},~
    \mathbf{o}^{i},~
    \mathbf{l} \setminus \{\mathbf{l}^{i}\},~
    \mathbf{o} \setminus \{\mathbf{o}^{i}\},~
    \mathbf{c}
\}$ and regard this process as a mapping function:

\begin{equation}
\uppi(\mathbf{s}^{i}) \mapsto \mathbf{a}^{i} 
\end{equation}
This mapping process is applied on each label at each timestamp.

Typically, a view management system should output actions to optimize the objectives.
We can define a reward function $r(\mathbf{s}^{i}, \mathbf{a}^{i})$ to evaluate how good the action $\mathbf{a}^{i}$ achieves the objectives given the input $\mathbf{s}^{i}$.
A good view management system for dynamic scenarios should maximize the reward in the long run. 
This can be formulated in an RL framework~\cite{sutton2018reinforcement}:

\begin{equation}\label{eq:pi}
    \arg \max_{\uppi} \{r(\mathbf{s}^{i}, \mathbf{a}^{i}) + \gamma V^*(\mathbf{s}^{i}, \mathbf{a}^{i}) | \uppi(\mathbf{s}^{i}) \mapsto \mathbf{a}^{i}\}
\end{equation}

\noindent
where $\gamma$ is a discount factor,
$V^*$ is the optimal value function that estimates the optimal cumulative rewards to the future after performing action $\mathbf{a}^{i}$ under state $\mathbf{s}^{i}$.
Simply put, 
\autoref{eq:pi} describes that
the optimal view management system $\uppi$ 
should generate actions to maximize not only the reward $r(\mathbf{s}^{i}, \mathbf{a}^{i})$ for the current state
but also the future cumulative rewards $\gamma V^*(\mathbf{s}^{i}, \mathbf{a}^{i})$.

Similar to previous works~\cite{DBLP:conf/icra/ChenLKA19, DBLP:conf/icra/ChenLEH17}, we aim to solve Eq.~\ref{eq:pi} using deep RL.
Specifically, we train a neural network to approximate an optimal mapping function $\uppi$.
To assist in the training, 
we also train a neural network to approximate the optimal value function $V^*$.

\para{The difference between RL- and optimization-based methods.}
Most of the existing systems for static objects formulate the view management problem as an optimization problem
that aims to optimize the reward for the \cmo{\emph{current state}}~\cite{DBLP:journals/cgf/BekosNN19}
(i.e., the first part of Eq.~\ref{eq:pi}, $\arg \max_{\uppi} \{r(\mathbf{s}^{i}, \mathbf{a}^{i}) \}$).
\cmo{
They excel when the scene is static, as the current state will not change over time.
However, in dynamic scenarios, where the state changes continuously, these methods tend to underperform 
because they cannot predict the \emph{future states} of objects.
Consequently, they may not provide the optimal or stable layout over time. 
In contrast, RL-based methods have the potential to overcome this issue by considering both current and future states.
}


\para{The challenges of placing labels in AR vs. other 3D dynamic scenarios.}
Placing labels on moving objects in AR environments presents a unique challenge compared to other similar 3D dynamic scenarios, such as VR or 3D games. 
In 3D games, the system can access the full states of objects 
and thus can \emph{plan} the best moving trajectories of the labels in advance.
In contrast, in AR environments, the system can only observe the current state of the objects and is uncertain about their future states, 
such as their velocities. 
Therefore, the system must have the capability to predict future states to place the labels properly. 

\section{\system{} Design}

\begin{figure}[h]
  \centering
  	\vspace{-2mm}
  \includegraphics[width=1\linewidth]{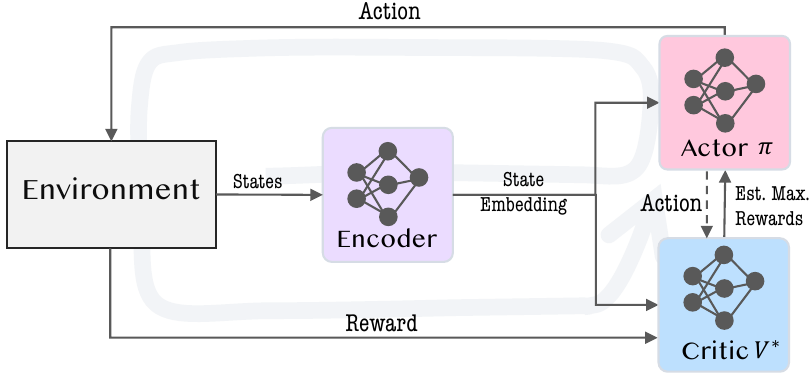}
  	\vspace{-2mm}
  \caption{Our RL-based method consists of three components: 
  an \emph{Encoder} that encodes the current state of the environment, an \emph{Actor} that generates actions for placing the labels, 
  and a \emph{Critic} that evaluates the generated actions and provides feedback to improve the actor based on the rewards obtained from the environment.}
  \vspace{-2mm}
  \label{fig:pipeline}
\end{figure}

\noindent
We introduce \system{},
which uses an \emph{Actor-Critic} framework~\cite{sutton2018reinforcement}
to manage the label placements of moving objects.
We chose the Actor-Critic framework since it has demonstrated state-of-the-art performance in various domains and benchmarks~\cite{Mnih2016, schulman2017proximal, DBLP:conf/icml/HaarnojaZAL18, DBLP:journals/corr/abs-2209-08480}.
It consists of three main components (\autoref{fig:pipeline}):

\begin{enumerate}[leftmargin=*]
    \item \textbf{Encoder for State Embedding} (Sec.\ref{sec:encoder}). 
    One of the main challenges of this work comes from state heterogeneity 
    (i.e., associating each label with the viewpoint and objects)
    and variability (i.e., encoding the relationship between each label and its neighbors, which can vary in number).
    We address the challenge through space transformations and a neural network with a self-attention mechanism.

    \item \textbf{Actor for Action Generation} (Sec.\ref{sec:actor}).
    We then use an Actor network to generate actions to place labels based on the state embedding.
    The challenge lies in designing an appropriate action space 
    that enables the model to converge during training 
    and generate effective actions to achieve objectives.
    Our action space consists of two accelerations (x- and z-)
    that control the movement of a label in an x-z square plane on top of its target object.

    \item \textbf{Critic for Reward Learning} (Sec.\ref{sec:critic}).
    To improve the effectiveness of the actions generated by the Actor network, 
    we introduce a Critic network that predicts the future cumulative reward for the actions 
    based on the current environment states and rewards.
    The environmental rewards are designed to incorporate the objectives of the label layout,
    such as avoiding occlusion, line intersections, and jittering.
    The Critic network is solely utilized to aid in training the Actor network and is not used during inference.

\end{enumerate}


\subsection{Encoder for State Embedding}
\label{sec:encoder}

\begin{figure}[h]
  \centering
  \vspace{-1mm}
  \includegraphics[width=1\linewidth]{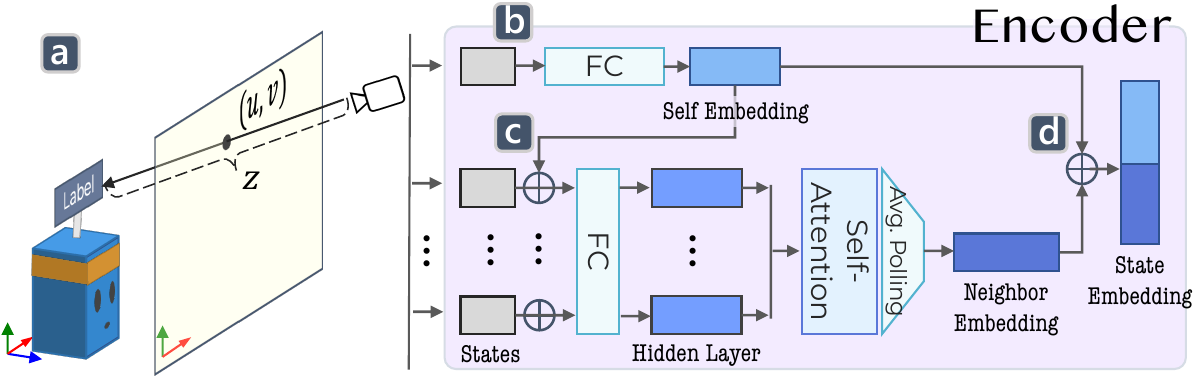}
  \caption{For each label or object, 
  the Encoder considers its state (e.g., position) relative to the camera and neighboring objects. 
  It then employs a neural network to embed the label or object's state into a high-dimensional vector.
  }
  \label{fig:encoder}
\end{figure}

\noindent
To train the neural network, 
it is necessary to convert the state information of labels, objects, and the viewpoint into a vector. 
We achieve this by 
a label-centered space transformation \cmo{and}
a neural network with a self-attention mechanism. 

\para{Space Transformation.}
To place a label $\mathbf{l}^{i}$,
the view management system needs to consider its state in relation to other labels, objects, and the viewpoint. 
We achieve this in two steps (\autoref{fig:encoder}a):

\begin{enumerate}[leftmargin=*]

\vspace{-1mm}
\item \emph{Neighbor Encoding.}
To encode the states of the label's neighbors (\emph{i.e.}, other objects and labels), 
we transfer their positions and velocities into the label's local space. 
This captures their relative positions and movements with respect to the label. 
Additionally, to distinguish between objects and labels, 
we append a binary value $w$ to each neighbor vector. 
A value of 1 indicates that a neighbor is an object, 
while a value of 0 indicates that the neighbor is a label.

\vspace{-1mm}
\item \emph{Viewpoint Encoding.}
To ensure that the label does not occlude other objects and labels, 
it is necessary to associate the viewpoint's state with the objects and labels. 
To encode the viewpoint's state,
we transfer all objects' and labels' positions to \emph{ray space} $\mathbb{R}^3_{ray}$,
in which a point $\mathbf{p}_{ray}$ is defined as $(u_{scn}, v_{scn}, z_{cam})$, 
where $(u_{scn}, v_{scn})$ is the point's position in screen space $\mathbb{R}^2_{scn}$
and the third coordinate $z_{cam}$ specifies the distance from the camera to the point's 3D position in $\mathbb{R}^3_{cam}$.
Such transformation can retain both the 2D and 3D information.

\end{enumerate}

The resulting states after the two steps are referred to as \emph{encoded states}.
Specifically,
the states of the label $\mathbf{l}^{i}_{ray}$, 
a neighbor object $\mathbf{o}^{j}_{ray}$, 
and a neighbor label $\mathbf{l}^{j}_{ray}$ are encoded as:
\begin{align*}
    \mathbf{l}^{i}_{ray} &= \{
\mathbf{p}_{\mathbf{l}, ray}^{i},~
\mathbf{v}_{\mathbf{l}}^{i},~
\mathbf{n}_{\mathbf{l}}^{i},~
\mathbf{p}_{\mathbf{o}, ray}^i - \mathbf{p}_{\mathbf{l}, ray}^{i}
\} \\
\mathbf{o}^{j}_{ray} &= \{
\mathbf{p}_{\mathbf{o},ray}^{j} - \mathbf{p}_{\mathbf{l}, ray}^{i},~ 
\mathbf{v}_{\mathbf{o}}^{j} - \mathbf{v}_{\mathbf{l}}^{i},~
\mathbf{n}_{\mathbf{o}}^{j},~
w: 1\}\\
\mathbf{l}^{j}_{ray} &= \{
\mathbf{p}_{\mathbf{l},ray}^{j} - \mathbf{p}_{\mathbf{l}, ray}^{i},~ 
\mathbf{v}_{\mathbf{l}}^{j} - \mathbf{v}_{\mathbf{l}}^{i},~ 
\mathbf{n}_{\mathbf{l}}^{j},~ 
\mathbf{p}_{\mathbf{o}, ray}^j - \mathbf{p}_{\mathbf{l}, ray}^{i},~ 
w: 0\}
\end{align*}


\vspace{1mm}
\para{State Embedding.}
Following the space transformation, we utilize a neural network to embed the resulting states into a high-dimensional space.
The network embeds the label $\mathbf{l}^{i}$ and its neighbors differently:

\begin{enumerate}[leftmargin=*]

\vspace{-1mm}
\item \emph{Self Embedding} (\autoref{fig:encoder}b). 
We first concatenate the encoded state of the label 
$\mathbf{l}^{i}_{ray}$ 
with the encoded state of its target object
$\mathbf{o}^{i}_{ray}$
and then embed them into a 128-length vector using a fully connected (FC) network.

\vspace{-1mm}
\item \emph{Neighbor Embedding} (\autoref{fig:encoder}c).
Building on previous work~\cite{DBLP:conf/icra/ChenLKA19},
we embedded the neighbors of the label $\mathbf{l}^{i}$ as \cmo{follows}:

For each of the label's neighbors (except for its target object),
we first concatenate its encoded state (e.g., $\mathbf{l}^{j}_{ray}$ or $\mathbf{o}^{j}_{ray}$) with the 128-length embedding vector obtained from the previous step.
Next, we embed the concatenated vector into a 128-length vector using an FC network.
Since the number of neighbors can vary across different scenes,
we use a \emph{self-attention mechanism}~\cite{DBLP:conf/icra/VemulaMO18} to address the variable number of inputs.
This network mechanism learns the relative importance of each neighbor as attention scores and 
uses the scores as weights to combine all the inputs into a fixed-size output.

\end{enumerate}

Finally, we concatenate the embedding of the label
and its neighbors into a single vector (\autoref{fig:encoder}d),
which contains rich information about the label, its neighbors, and the viewpoint of the current moment.

\subsection{Actor for Action Generating}
\label{sec:actor}

\begin{figure}[h]
  \centering
  \includegraphics[width=1\linewidth]{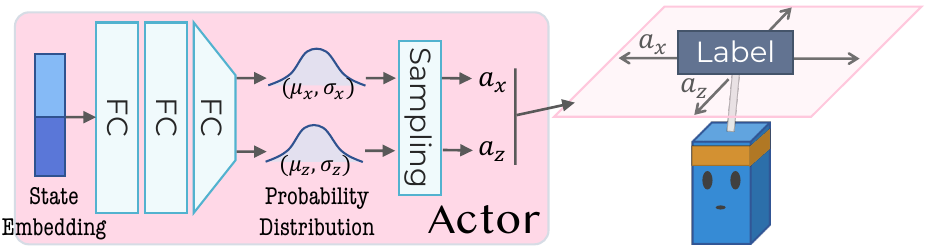}
  \caption{An Actor network generates actions based on the state embedding to place the label. 
The label's movement is constrained to a two-dimensional x-z square plane on top of its target object, and the available actions consist of x- and z- accelerations for the label.
  }
  \label{fig:actor}
\end{figure}

\noindent
The system should generate actions to place the labels.
To achieve this, we first define an action space that contains all possible actions,
and then use an Actor network to learn a policy (i.e., $\pi$) that generates actions from this space.

\para{Action Space.}
The action space plays a central role in successfully training a model:
a large action space can make the model fail to converge
while a small one may not be able to generate effective actions that fulfill the objectives (e.g., occlusion free, stable movements).
\cmo{Inspired by Tatzgern et al.~\cite{DBLP:conf/vr/TatzgernKGS14}, 
we first constrain the label's movement to an x-z square plane on top of its target object (\autoref{fig:actor}, right)}.
We choose this plane to ensure consistency in the degrees of freedom of the labels and objects, 
as we assume that the objects can only move in the x and z dimensions.
We then define the actions generated by the model as x- and z- accelerations (i.e., $a_x$ and $a_z$) that control the label's movement to ensure stability. 
In other words, the action space consists of 
two accelerations 
that control the label to move in an x-z square plane on top of its target object.
In our study, we also explored another action space in which the model generated the x and z velocities or positions of the labels,
but this led to jittering movements due to the non-deterministic nature of the network output (see below).


\para{Action Generation.}
We use a three-layer FC network to learn a mapping function $\pi$ that generates actions.
The network inputs the embedding vector of label $\mathbf{l}^{i}$  
and outputs the means and standard deviations of two normal distributions.
These two distributions serve as probability distributions to sample the accelerations for the x and z directions of the label (\autoref{fig:actor}, left).
This stochastic policy encourages the network to explore different actions during training, rather than only selecting the same action every time, which can help reduce the possibility of getting stuck in a locally optimal solution
and has been widely used in RL to improve performance~\cite{sutton2018reinforcement}.


\begin{figure*}[b!]
    \setcounter{figure}{7}
  \centering
  \includegraphics[width=\linewidth]{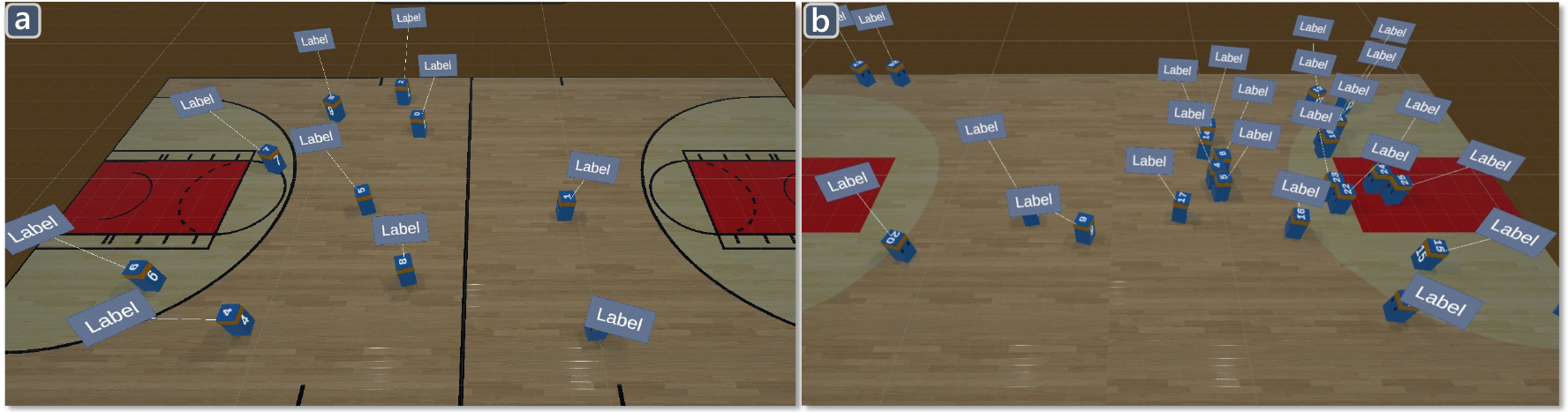}
  \vspace{-2mm}
  \caption{
  a) the \NBA{} dataset, in which each scene contains 10 fast-moving objects.
  b) the \STU{} dataset, in which the number of objects is dynamic. Objects in the \STU{} dataset move slower than those in the \NBA{} dataset.
}
  \label{fig:datasets}
\end{figure*}

\begin{figure}[th]
\setcounter{figure}{6}
  \centering
  \includegraphics[width=0.5\linewidth]{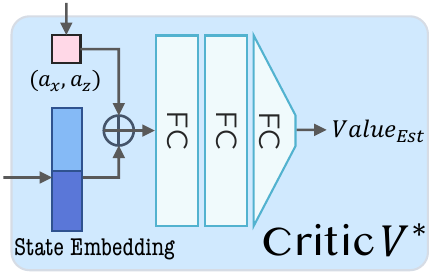}
  \caption{The Critic network estimates the future accumulated reward for the actions generated by the Actor, 
  based on the current state of the environment. 
  The estimated reward is then used to refine the Actor.}
  \vspace{-3mm}
  \label{fig:critic}
\end{figure}

\subsection{Critic for Reward Learning}
\label{sec:critic}

\noindent
To improve the Actor network's ability in generating effective actions, 
we designed rewards to reflect the feedback from the environment, 
incorporating the objectives of the label layout, 
such as avoiding occlusion, line intersections, and jittering.
However, these rewards only provide feedback for the current moment. 
To improve long-term performance, we use a Critic network ($V^*$).
The Critic learns from the rewards, actions, and states to predict the future cumulative reward of a given action on the current state, 
guiding the Actor network.


\para{Reward Design.}
In line with previous work~\cite{DBLP:journals/cgf/BekosNN19},
we aim for the labels to move without occlusion, line intersection, and jittering.
Thus, we design the reward from the environment
as $r(\mathbf{s}^{i}, \mathbf{a}^{i}) = r_{occ} + r_{int} + r_{a_{x,y}} $,
which awards achieving each objective while penalizing any failures.
Specifically, each term is defined as follows:

\vspace{-4mm}
\begin{align*}
r_{occ} &=
  \begin{cases}
  -0.1n_{occ}, & \text{if } n_{occ} > 0 \\ 
  0.1, &\text{otherwise}
\end{cases}\\
r_{int} &=
  \begin{cases}
  -0.1n_{int}, &\text{if } n_{int} > 0 \\ 
  0.1, &\text{otherwise}
  \end{cases}\\
r_{a_{x,y}} &=
  \begin{cases}
  0.001, &\text{if } \abs{a_x} \le \max_{a_x} \text{ and } \abs{a_y} \le \max_{a_y} \\ 
  -0.001, &\text{otherwise}
  \end{cases}
\vspace{-2mm}
\end{align*}

\noindent
where $n_{occ}$ represents the number of objects occluded by a label, 
$n_{int}$ represents the number of lines intersected by the label's leader line,
and $\max_{a_x}$ and $\max_{a_z}$ represent the predefined maximum accelerations on x- and z- dimensions. 
We chose the number of rewards (e.g., 0.1, 0.001) 
based on prior works~\cite{DBLP:conf/icra/ChenLKA19, juliani2018unity}
and empirical evaluations.

\cmo{
Reward engineering~\cite{DBLP:conf/aaaiss/Dewey14} is an iterative process, similar to fine-tuning hyperparameters. 
We arrived at our current design through extensive exploration of reward designs, including variations in positive/negative reward values, 
reward magnitudes, 
incorporation of occlusion area, 
and measuring label movement distances to assess jittering. 
None of them demonstrated the same level of performance as our current design.
}

\para{Cumulative Rewards Prediction.}
To help the Actor network generate actions that maximize both the reward for the current state and the future cumulative rewards,
we used three FC layers to approximate the optimal value function $V^*$. 
This network predicts the future cumulative reward based on the current state and the action generated by the Actor.
The predicted reward is then used to calculate the loss to train the Actor. 
Note that the Critic network is only used as an assistant during training and is not necessary for inference~\cite{sutton2018reinforcement}.

\section{Dataset and Network Training}
To train and evaluate \system{},
we collected two human movement datasets
to simulate real-world scenarios with moving objects (Sec.~\ref{sec:data}).
We first introduce the training process (Sec.~\ref{sec:network_training})
and then report the quantitative evaluation results (Sec.~\ref{sec:computational_eval}).

\subsection{Dynamic Environments Simulation}
\label{sec:data}


\para{Real-world Human Movement Datasets.}
We used two popular human movement datasets to simulate real-world environments with moving objects.
The first dataset, \emph{\NBA{}}~\cite{sportvu},
consists of the trajectories of 10 players on the court
during the first quarter (720 seconds in total) of a well-known NBA game~\footnote{Golden State Warriors and Cleveland Cavaliers on Dec 25th, 2015}.
The second dataset, \emph{\STU{}}~\cite{DBLP:journals/cgf/LernerCL07},
records the trajectories of students in an open campus environment over a period of 400 seconds.
Unlike the \NBA{} dataset,
the number of objects over time in the \STU{} dataset is dynamic,
as some students enter or leave the campus environment during the period.
Both datasets represent the trajectory of an object (i.e., player or student) as a list of 2D positions taken at an interval of 0.04 seconds.
In total, 
the \NBA{} and \STU{} datasets contain over 180K 
and 160K positions, respectively.

\begin{table}[h]
	\centering
    \small
    	\vspace{-2mm}
 	\caption{Statistics of scenes in the two datasets.}
	\label{table:scenes}
	\vspace{-2mm}
	\begin{tabular}{lccc}
		\toprule
		 Dataset & Avg. Max. $\#$Objects & Avg. Speed (m/s) & Avg. Moving Distance (m) \\ \midrule
		 \NBA{}         & 10  &  1.88  & 28.19   \\ 
		 \STU{}         & 20   &  1.29    & 9.82      \\ 
		\bottomrule
	\end{tabular}
 	\vspace{-2mm}
\end{table}

We preprocessed the \NBA{} and \STU{} datasets by dividing them into small scenes lasting 15 seconds each, so that there is no overlap between objects' trajectories in different scenes.
We excluded scenes that are less than 15 seconds long. 
For the \NBA{} dataset, 
we further removed scenes where the game stops due to events such as fouls or substitutions.
Ultimately, we gathered 26 scenes for each dataset.
Table.~\ref{table:scenes} summarizes the statistics of the scenes for each dataset.
The \STU{} dataset is generally more challenging due to the larger and dynamic number of objects.

\para{Simulated Environments in Unity.}
We used Unity to simulate the scenes in the two datasets. 
Specifically, 
we created 3D cubes to represent the objects in the scenes and updated the cubes' positions every 0.04 seconds to follow the objects' trajectories.
An example scene for each dataset is shown in \autoref{fig:datasets}a and b.

\subsection{Network Training}
\label{sec:network_training}


We followed machine learning conventions and randomly split the scenes of each dataset into two sets: 80\% of the scenes for training and 20\% of the scenes for testing. 
This resulted in 20 scenes for training and 6 scenes for testing in each dataset.


\para{Loss Function.}
Our method adopts an actor-critic framework~\cite{sutton2018reinforcement}, which usually involves two loss terms:

\begin{equation}
    \label{eq:policy_loss}
    \mathcal{L}_\pi = -\frac{1}{nT} \sum_{i=0}^{n} \sum_{t=0}^{T} \log\pi (\mathbf{a}^{i,t} | \mathbf{s}^{i, t}) A_{i, t}
\end{equation}
\begin{equation}
    \label{eq:value_loss}
    \mathcal{L}_V = -\frac{1}{nT} \sum_{i=0}^{n} \sum_{t=0}^{T} (V^*(\mathbf{s}^{i, t}, \mathbf{a}^{i,t}) - R_t)^2
\end{equation}

\noindent
where $T$ represents the maximum time step of a scene, 
$n$ is the number of labels.
Eq.~\ref{eq:policy_loss} penalizes the actor's actions 
that have a high probability of occurring but result in low advantages 
$A_{i, t} = 
r(\mathbf{s}^{i, t}, \mathbf{a}^{i, t}) + 
\gamma V^*(\mathbf{s}^{i, t}, \mathbf{a}^{i,t})
- V^*(\mathbf{s}^{i, t-1}, \mathbf{a}^{i,t-1})
$,
which is an estimation of the future cumulative rewards with bias subtraction
for the actions in the current state.
Eq.~\ref{eq:value_loss} trains the critic's estimation to match 
the future cumulative rewards, $R_t = \sum_{t' = t}^{T} \gamma^{t' - t} r(\mathbf{s}^{i, t'}, \mathbf{a}^{i, t'})$.
In our implementation, we use PPO~\cite{schulman2017proximal},
a state-of-the-art actor-critic algorithm,
which involves several advanced techniques to calculate Eq.~\ref{eq:policy_loss},
such as using generalized advantage estimation~\cite{schulman2015high} to compute $A_{i, t}$,
differential entropy loss to encourage exploration and avoid premature convergence,
and clipped surrogate objective to stabilize the training process.
We refer the readers to Schulman et al.~\cite{schulman2017proximal} for more technical details.

\para{Hyper Parameters.}
Two hyper parameters play important roles in our method, namely, \emph{maxAcc} and \emph{numAgent}:
\begin{itemize}
    \item \emph{maxAcc} controls the \cmo{absolute value of the} maximum acceleration the actor can apply for the labels. 
    A larger \emph{maxAcc} can  
    reduce occlusions
    but increases the jittering of the labels. 
    Based on the mean moving speed of objects, 
    we set the \emph{maxAcc} for the \NBA{} and \STU{} datasets to 3$m/s^2$ and 2$m/s^2$, respectively.
    
    \item \emph{numAgent} specifies the number of labels to be controlled by the model.
    Ideally, the model should control all the labels in a scene. 
    However, a larger \emph{numAgent} can make training more difficult.
    To address this issue, we employ a technique called Curriculum Learning~\cite{narvekar2020curriculum},
    in which we gradually increase the \emph{numAgent} during training.
    We increase the \emph{numAgent} from 2 to 10 (stepsize: 2) and 4 to 20 (stepsize: 4) for the \NBA{} and \STU{} datasets, respectively.
\end{itemize}

\para{Implementations Details.}
We use ML-Agent~\cite{juliani2018unity}, 
a Unity-based reinforcement learning toolkit,
to implement our system.
ML-Agent provides an implementation of the RL networks using PyTorch~\cite{paszke2019pytorch},
as well as components
to connect the Unity environments with the PyTorch backends
and manage the training.
To accelerate the training, 
we also leverage the concurrent Unity instances feature provided by ML-Agent.
Accordingly, we choose 204,800 and 1,024 for the buffer\_size and batch\_size
to ensure the models collect rich enough experiences from the simulated environments.
The learning rate is initialized as 3e-4 and decays linearly over training.
Each training process contains 20 million steps,
\cmo{leading to about 1.3 million episodes (each episode lasts 150 steps)}.
The training process for the models was conducted on servers equipped with NVIDIA V100 graphics cards and 6 vCPUs, 
and took approximately 5 hours to complete.

\section{Computational Experiments}
\label{sec:computational_eval}

To evaluate our method,
we first examined the effectiveness of the training process (Sec.~\ref{sec:under_over}),
then qualitatively inspected the ability of the value network to predict future rewards (Sec.~\ref{sec:value_heatmap}),
compared our method with two baselines to assess its performance in handling occlusions, line intersections, and label jittering (Sec.~\ref{sec:under_over}),
and finally discuss some typical behaviors of the model (Sec.~\ref{sec:cases}).

\subsection{Rewards Over the Training}
\label{sec:under_over}

\begin{figure}[h]
  \centering
    \setcounter{figure}{8}
  \includegraphics[width=1.05\linewidth]{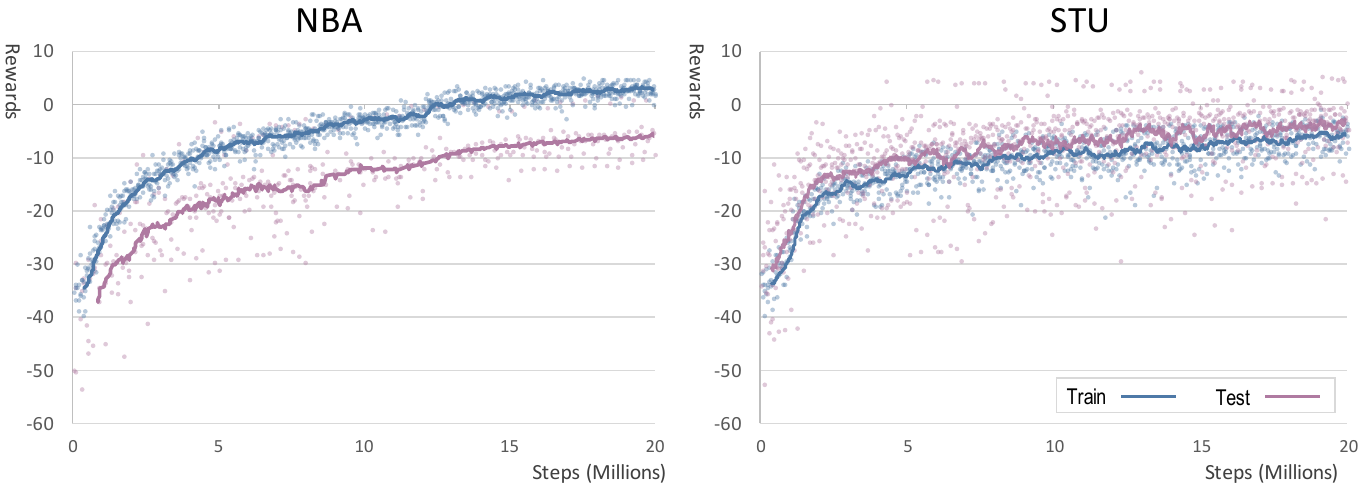}
  \caption{Accumulated rewards per epoch in training (blue) and testing (purple) scenes for \NBA{} (left) and \STU{} (right) datasets.}
  \label{fig:rewards}
\end{figure}

\noindent
Figure~\ref{fig:rewards} demonstrates the rewards obtained by the RL model in the training (blue) and testing (purple) scenes during the training process.
The plot indicates that both the rewards 
increase over the training process,
indicating that the model is effectively learning and improving. 
\cmo{
The negative rewards in our approach stem from the strict reward definition, where occlusion is considered to occur whenever two labels overlap. 
However, from a human perspective, these occlusions can often be overlooked. 
Despite the presence of negative rewards, the scenes may still appear occlusion-free to the audience.
}

\subsection{Value Heatmap Learned by the Model}
\label{sec:value_heatmap}

\begin{figure}[h]
  \centering
  \includegraphics[width=1\linewidth]{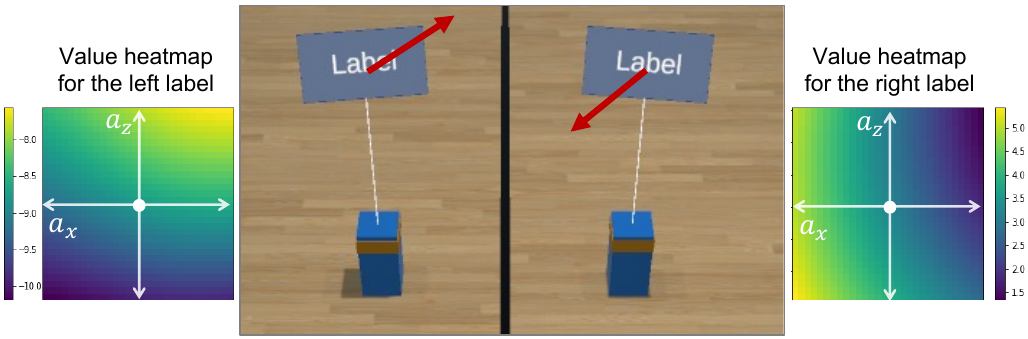}
  \caption{Two value heatmaps generated by our model in a scenario where two objects move towards each other. 
    Each heatmap represents the estimated future rewards of actions in the corresponding object's action space.
  Brighter colors indicate areas where the network considers a higher reward for moving, 
  while darker colors indicate areas where the network does not prefer movement.
  }
  \label{fig:value_map}
\end{figure}

\noindent
We conducted a qualitative inspection of the value network to assess its ability to estimate the accumulated rewards. 
We set up a scenario where two objects move towards each other and divided the action space into a 30x30 grid.
Then, we estimated the rewards for different grids across the action space using the value network and visualized them as value heatmaps. 
Figure~\ref{fig:value_map} displays examples of these heatmaps. 
The network suggests that the left label can achieve the maximum reward if it moves to the top right, 
while the right label should move to the bottom left. 
As a result, both objects avoid occluding each other. 
These examples demonstrate that the value network has learned to meaningfully estimate the consequences of different actions.

\begin{figure*}[b]
  \centering
  \includegraphics[width=\linewidth]{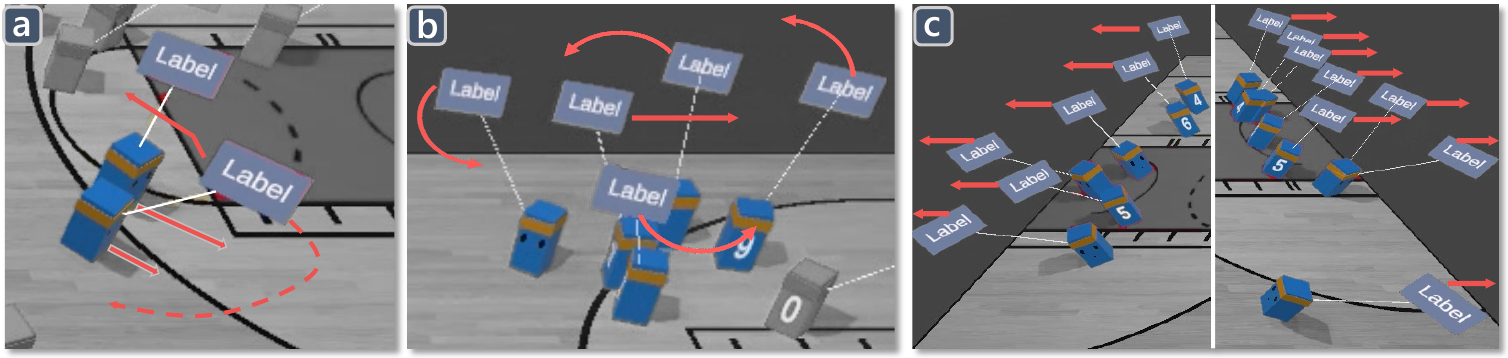}
  \caption{
Examples of the model's capabilities in controlling labels to avoid occlusion and line intersections. 
a) The model minimizes label movement by utilizing the small gap between an object and its label to avoid occlusion.
b) The model controls labels to move collaboratively in a circular path around objects, eliminating any line intersections
c) The model strategically places labels outside the court to reduce occlusion.
  }
  \label{fig:cases}
\end{figure*}

\subsection{Performance in Achieving Objectives}
\label{sec:comp_eval}

\noindent
We compared three different view management methods:
1) \textbf{\No{}} view management as the baseline,
where a label is fixed on top of its target object, 
2) a \textbf{\Force{}}-based method,
and 3) \textbf{\Ours{}}, the RL-based method.
Since there is no existing view management system for labels of moving objects, 
we chose to
implement and adapt \Force{} based on Plane-based Hedgehog Labeling~\cite{DBLP:conf/vr/TatzgernKGS14},
a state-of-the-art method for managing 3D labels for static objects,
under the guidance of their authors.
Note that our comparison only intends to use \Force{} as a reference instead of showing our method is better than Hedgehog Labeling~\cite{DBLP:conf/vr/TatzgernKGS14}.

Ideally, the view management system should reduce occlusions, line intersections, and the jittering of labels.
We measure three metrics for each label to assess these three objectives:

\begin{itemize}
    \item \emph{OCC} measures 
    the average number of objects and labels occluded by a label in each time step.
    A smaller OCC value is better.

    \item \emph{INT} measures how many leader lines are intersected by a label's leader line on average in each time step, with a lower score indicating better performance.

    \item \emph{DIST} measures the average extra moving distance of each label compared to its target object.
    DIST approximates the degree of jittering of the labels, and a smaller value is better. 

\end{itemize}

\begin{table}[th]
	\centering
 	\caption{Performance of three methods in reducing occlusion, line intersections, and extra movement distance of the labels.}
 \vspace{-2mm}
	\label{table:error_rate}
	\begin{tabular}{lcccccc}
		\multirow{2}{0cm}{} & \multicolumn{3}{c}{\textbf{NBA}} & \multicolumn{3}{c}{\textbf{STU}} \\
		                        \cmidrule(l{2pt}r{2pt}){2-4} \cmidrule(l{2pt}r{2pt}){5-7}
		              & OCC & INT & DIST & OCC & INT & DIST \\ \midrule
		\textbf{No}    & 0.15 & 0 & 0 & 0.18 & 0 & 0 \\ \hline
		\textbf{Force} & 0.05 & 0.04 & +11.29 & 0.07 & 0.06 & +11.46 \\ 
		\textbf{Ours} & \textbf{0.04} & \textbf{0.02} & \textbf{+7.34} & \textbf{0.06} & \textbf{0.02} & \textbf{+2.58} \\ 
		\bottomrule
	\end{tabular}
\vspace{-2mm}

\end{table}

Table~\ref{table:error_rate} presents the results of the three methods on the three metrics.
For OCC, without using a view management system, 
each label occludes 0.15 and 0.18 label and objects per time step in the \NBA{} or \STU{} datasets, respectively.
Both \Ours{} and \Force{} succeeded in reducing label occlusions by over two-thirds compared to the \cmo{\No{} baseline}.
As for INT, without view management, it is zero since all the leader lines are perpendicular to the ground.
Table~\ref{table:error_rate} shows that the intersection introduced by \Ours{} is half that of \Force{}.
Similarly, 
DIST should be zero without view management since the moving distance of a label should be the same as its target object.
Table~\ref{table:error_rate} indicates that \Ours{} has much smaller DIST than \Force{}, 
with \Ours{} being almost one-fifth of \Force{} in crowded \STU{} scenarios.
In general, these results show that \Ours{} can reduce occlusions with fewer line intersections and more stable movements 
than \Force{}.


\subsection{Examples}
\label{sec:cases}

We manually observed the model's control of the labels in all scenes to understand how it achieves the objectives.
Here, we discuss some typical behaviors of the model.
Video examples are also provided in the supplemental materials to illustrate these behaviors.


\para{Avoiding occlusions through minimal movement.}
Figure~\ref{fig:cases}a shows an example of the model's capacity to avoid occlusion by moving labels minimally.
In this example, two objects are moving forward, with the rear object moving faster.
To prevent occlusion of the rear object's label, 
the model controls the front object's label to move backward, 
using the small gap between the rear object and its label, 
rather than taking an easier but longer distance route (as indicated by the dashed arrow). 
This demonstrates the model's high controllability, with the task being completed within 0.2 seconds.

\para{Avoiding line interactions through collaborative label movement.}
Figure~\ref{fig:cases}b illustrates how the model controls the labels to move collaboratively to avoid line intersections in a complex movement scenario.
The labels on the outside move in a circular path around the objects, similar to driving around a roundabout, while the label of the center object moves straight forward to the right.
This strategy effectively eliminates any line intersections.
This example demonstrates the model's ability to observe the states of a label's neighbors 
and adjust the label movement based on their relationships, 
highlighting its capability to handle complex situations.

\para{Reducing Occlusion through strategic label placements.}
In \autoref{fig:cases}c,
the model is shown strategically placing the labels outside the court 
to reduce the likelihood of occlusion.
When the objects move to the left side of the court, 
the model moves the labels to the empty space on the left, 
and vice versa when the objects are on the right. 
This behavior suggests that the model has learned to associate the empty space outside the court with safety based on the movement of objects and environmental feedback.

\section{User Study}
We conducted a controlled user study to evaluate if \system{}
can help users perform visual tasks in dynamic scenes.

\subsection{Experiment Settings}

\begin{figure}[h]
  \centering
\vspace{-2mm}
  \includegraphics[width=1\linewidth]{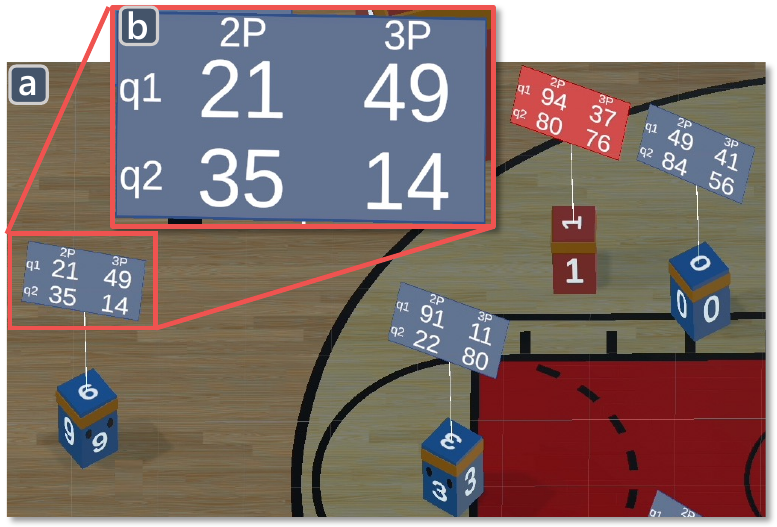}
  \vspace{-2mm}
  \caption{
      a) A scene example of a task in which the objects are divided into blue and red teams.
      b) A table on a label, where the rows and columns represent quarters and points, respectively.
  }
  \vspace{-3mm}
  \label{fig:study_example}
\end{figure}

\para{Participants and Apparatus.}
We recruited 18 participants (P1-P18; M = 8, F = 8; Age: 20 - 40)
through university mailing lists and forums.
Nine had no experience with AR/VR devices, seven had less than 1 year of experience, and one had 1-3 years of experience.
Only one participant had no experience with data visualizations, while others had 0 to 5+ years of experience.
The study was conducted in a quiet lab room using a standalone wireless Oculus Quest 2 VR headset, allowing participants to move freely in space without being limited by headset cables. The study took approximately 60 minutes, and each participant received a \$20 gift card as compensation.
 
\para{Conditions.}
We compared the three view management methods mentioned in Sec.\ref{sec:comp_eval}, i.e., \No{}, \Force{}, and \Ours{}.

\para{Datasets.}
We reused the \NBA{} and \STU{} datasets for the user study.
We used the scenes in the training set for practice trials
and the scenes in the testing set for the actual study.
In each scene, the objects are divided into blue and red teams.
Each scene contains objects with a unique ID displayed on their bodies (\autoref{fig:study_example}a), divided into blue and red teams.
Each object has a label attached to it, showing a static data table (\autoref{fig:study_example}b), 
in which the rows and columns represent quarters and points, respectively.
Each scene has a duration of 15 seconds.

\para{Tasks.}
Following the previous work~\cite{DBLP:journals/tvcg/LinYBP23} on AR labels,
we designed three basic visual search tasks~\cite{DBLP:journals/tvcg/BrehmerM13}:
\begin{itemize}
    \item \IDE{}:
    This task asks the participants to identify the points a specific player scored in a specific quarter.
    The participants need to answer a question in the format of \emph{``How many \underline{[2, 3]} points has player \underline{[X]} got in the quarter \underline{[1, 2]}?''}, in which the underlined numbers are randomized across participants.

    \item \COM{}:
    This task asks the participants to compare the points the players in a specific team got in a specific quarter.
    The participants need to answer a question in the format of \emph{``In the \underline{[blue, red]} team, who got the most \underline{[2, 3]} points in the quarter \underline{[1, 2]}?''}, in which the underlined numbers are randomized across participants.

    \item \SUM{}:
    This task asks the participants to summarize the points a team scored in a specific quarter.
    The participants need to answer a question in the format of \emph{``Overall, which team got more \underline{[2, 3]} points in the quarter \underline{[1, 2]}?''}, in which the underlined numbers are randomized across participants.
\end{itemize}

\para{Study Design.}
We used the following full-factorial within-subject study design with Latin square-randomized order of the techniques:

\[
\def\arraystretch{1.2}
\begin{array}{c c l}
  & 18 & \text{participants} \\
\times & 3 & \text{techniques: \No{}, \Force{}, \Ours{}} \\
\times & 2 & \text{datasets: \NBA{}, \STU{}} \\
\times & 3 & \text{tasks: \IDE{}, \COM{}, \SUM{}}\\
\times & 2 & \text{timed repetition} \\
\cline{1-3}
  & 648 & \text{total trials (36 per participant)} \\
\end{array}
\]

Participants were split evenly between the 3 Latin square-randomized technique orders.
The orders of the datasets and tasks were kept consistent across all participants.
Each task had randomized data on the labels and was repeated in different scenes.

\para{Procedure.}
For each study, we first introduced the motivation, tasks, and general procedure, followed by obtaining consent (10mins).
Before the actual tasks, 
participants completed a training session (10mins) to become familiar with the tasks, user interfaces, and to adjust the device until they were comfortable.
The training session consisted of 18  = 3 (techniques) x 3 (tasks) x 2 (datasets) trials.
Once the participant was confident, 
the experimenter proceeded to the actual tasks, 
which consisted of two sessions for the \NBA{} and \STU{} datasets, respecitvely.
Each session (10mins) included 18 = 3 (techniques) x 3 (tasks) x 2 (datasets) x 2 (repetition) trials.
For each trial, the participants were instructed to finish the task \emph{as fast and accurately as possible} 
and to click a button to stop the timer once they were confident to answer.
The scene in each trial looped until the participants clicked the button, after which it disappeared.
After speaking their answer aloud, participants were asked to rate their mental load for the task.
Participants were allowed to take a break between each session. 
At the end of the study, participants provided subjective feedback on each technique (5mins).

\para{Measures.}
For each trial, 
we recorded the completion time (in seconds, from start to button click), 
accuracy (true or false), 
and the user's perceived mental load (on a 1-7 Likert scale).

\subsection{Quantitative Results}
We analyzed the statistical differences between the three methods in accuracy, completion time, and mental load.
Overall, both \Ours{} and \Force{} are significantly better than \No{} in all three measures
and \Ours{} slightly better than \Force{} in completion time and mental load.


{\setlength{\tabcolsep}{0em}
\begin{table}[th]
	\centering
        \small
        \caption{Accuracy per method per task on the two datasets.}
 	\label{table:error_rate}
	\vspace{-2mm}
	\begin{tabular}{p{8mm} >{\centering}p{13mm} >{\centering}p{12mm} >{\centering}p{14mm} >{\centering}p{13mm} >{\centering}p{12mm} >{\centering\arraybackslash}p{14mm}}
		\multirow{2}{0cm}{} & \multicolumn{3}{c}{\textbf{NBA}} & \multicolumn{3}{c}{\textbf{STU}} \\
		                        \cmidrule(l{2pt}r{2pt}){2-4} \cmidrule(l{2pt}r{2pt}){5-7}
		           & \IDE{} & \COM{} & \SUM{} & \IDE{} & \COM{} & \SUM{} \\ \midrule
		\No{}    & \textbf{100\%} & 83\% & 97\% & 50\% & 17\% & \textbf{100\%}  \\
		\Force{} & 97\% & 94\% & \textbf{100\%} & 92\% & 67\% & 97\% \\ 
		\Ours{} & 97\% & \textbf{97\%} & 97\% & \textbf{97\%} & \textbf{75\%} & \textbf{100\%} \\ 
		\bottomrule
	\end{tabular}
        \vspace{-2mm}

\end{table}
}

\para{Accuracy.}
Table~\ref{table:error_rate} shows the accuracy of participants in different tasks using the three methods.
The mean accuracy, in percentage, for \NBA{} were: 
$\text{\No{}} = 93\%$ ($\sigma = 17\%$), 
$\text{\Force{}} = 97\%$ ($\sigma = 12\%$), 
and $\text{\Ours{}} = 97\%$ ({$\sigma = 12\%$}).
For \STU{}, the mean accuracy were:
$\text{\No{}} = 56\%$ ($\sigma = 44\%$), 
$\text{\Force{}} = 85\%$ ($\sigma = 27\%$), 
and $\text{\Ours{}} = 91\%$ ($\sigma = 22\%$).
A Shapiro-Wilk test reveals that the accuracy did not follow a normal distribution.
Thus, we used a Friedman test with a null hypothesis that participants performed equally correctly with each method,
which showed significant differences in the \IDE{} ($p=.0001$) and \COM{} ($p<0.0001$) tasks on the \STU{} dataset.
By further performing a Nemenyi post-hoc test,
we found that participants performed significantly better in the \IDE{} tasks with \Ours{} ($p=.0099$) or \Force{} ($p=.0265$) than with \No{}.
Similarly, participants performed significantly better in the \COM{} task with \Ours{} ($p=.001$) or \Force{} ($p=.0018$) than with \No{}.


\begin{figure}[h]
  \centering
  \includegraphics[width=1\linewidth]{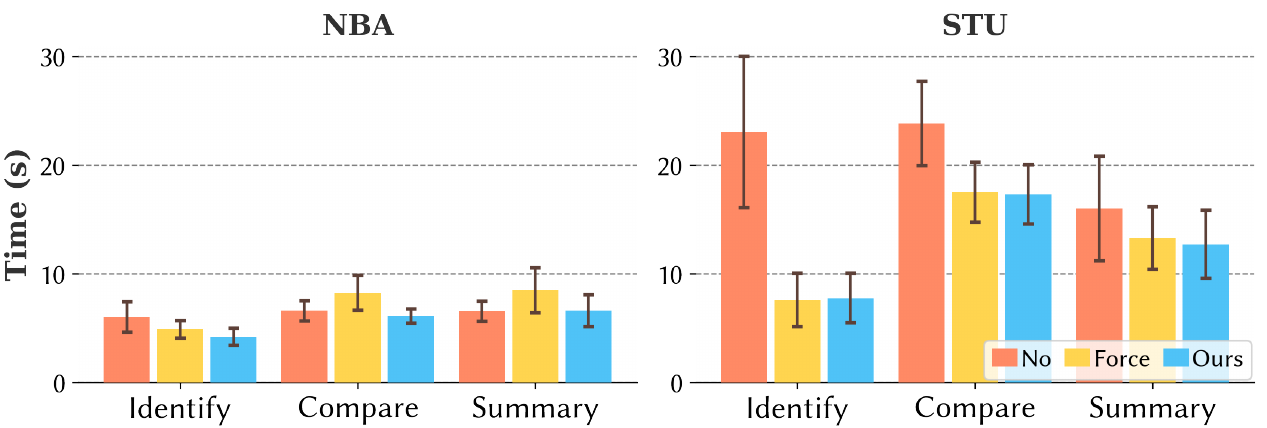}
  \vspace{-2mm}
  \caption{Completion time per method per task on the two datasets.}
\vspace{-2mm}
  \label{fig:time}
\end{figure}

\vspace{-1mm}
\para{Completion Time.}
Figure~\ref{fig:time} presents the completion time of participants in different tasks using different methods with 95\% confidence intervals (CIs).
The respective mean times in seconds are $\text{\No{}} = 6.38$ ($\sigma = 3.25$), 
$\text{\Force{}} = 7.20$ ($\sigma = 4.94$), 
and $\text{\Ours{}} = 5.63$ ({$\sigma = 3.20$}) on \NBA{},
and $\text{\No{}} = 20.97$ ($\sigma = 16.15$), 
$\text{\Force{}} = 12.79$ ($\sigma = 8.91$), 
and $\text{\Ours{}} = 12.60$ ($\sigma = 8.95$) on \STU{}.
Through a Shapiro-Wilk test, we confirmed that the completion times in each condition did not follow a normal distribution.
Using a Friedman test with a null hypothesis that participants performed equally fast with each method,
we found significant differences in the \COM{} ($p=.0387$) task on \NBA{}, 
and in both the \IDE{} ($p<0.0001$) and \COM{} ($p=.0293$) tasks on \STU{}.
A Nemenyi post-hoc test revealed that participants completed the \COM{} task significantly faster 
with \Ours{} than \Force{} ($p=.0355$) on \NBA{},
and significantly faster with \Ours{} than \No{} in the \IDE{} ($p=.001$) and \COM{} ($p=.0497$) tasks on \STU{}.
Participants also completed the \IDE{} ($p=.001$) and \COM{} ($p=.0483$) tasks significantly faster with \Force{} than \No{} on \STU{}.


\begin{figure}[h]
  \centering
  \includegraphics[width=1\linewidth]{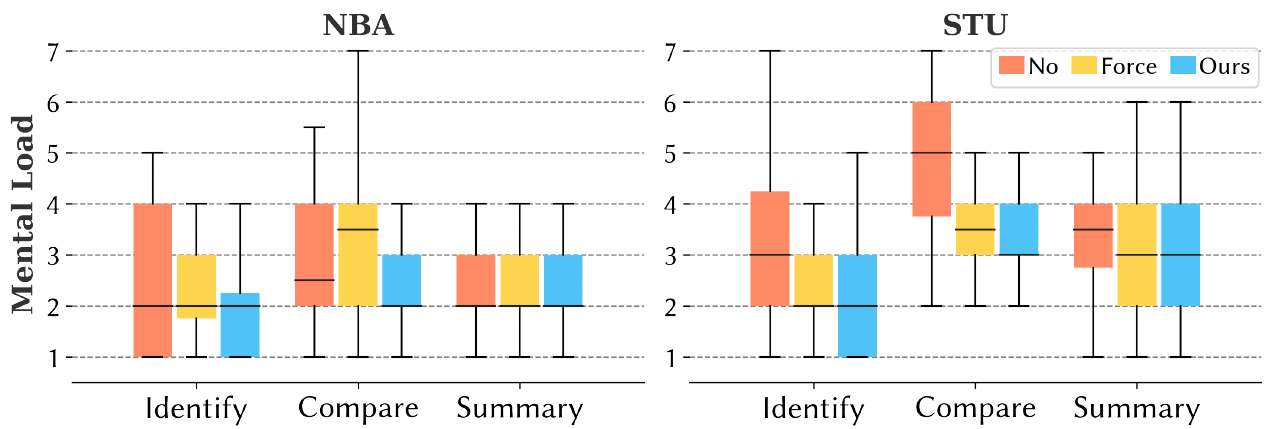}
  \vspace{-2mm}
  \caption{Mental load per method per task on the two datasets.}
  \vspace{-2mm}
  \label{fig:mental_load}
\end{figure}

\vspace{-1mm}
\para{Mental Load.}
Figure~\ref{fig:mental_load} shows the participants' subjective ratings on the mental load of using different methods in different tasks after each trial,
where 1 means low and 7 means high mental load.
The respective mean ratings are 
$\text{\No{}} = 2.55$ ($\sigma = 1.1$), 
$\text{\Force{}} = 2.74$ ($\sigma = 1.29$), 
and $\text{\Ours{}} = 2.36$ ({$\sigma = 1.08$}) on \NBA{},
and $\text{\No{}} = 3.83$ ($\sigma = 1.59$), 
$\text{\Force{}} = 3.02$ ($\sigma = 1.27$), 
and $\text{\Ours{}} = 3.01$ ($\sigma = 1.20$) on \STU{}.
Since a Shapiro-Wilk test shown that the ratings did not follow a normal distribution,
we used a Friedman test with a null hypothesis that the participants had the same mental load with the three methods.
Findings revealed significant differences in the \COM{} ($p=.0071$) task on \NBA{} 
and in both the \IDE{} ($p=.0004$) and \COM{} ($p < .0001$) tasks on \STU{}.
After performing a Nemenyi post-hoc test,
we found that participants reported significantly higher mental load with \Force{} than with \Ours{} in the \COM{} ($p=.0483$) task on \NBA{},
and significantly higher mental load with \No{} than with \Ours{} in both the \IDE{} ($p=.013$) and \COM{} ($p=.009$) tasks on \STU{}.
Furthermore, participants reported significantly higher mental load with \No{} than with \Force{} in both the \IDE{} ($p=.0075$) and \COM{} ($p=.001$) tasks on \STU{}.


\subsection{Qualitative Feedback}

We also collected qualitative feedback from the participants after the study.
Participants were asked to comment on the pros and cons of each method.
Overall, participants thought that \Ours{} is the most promising method as it achieves occlusion-free and stable labels.
\begin{itemize}
    \item \No{}.
    Participants provided mixed feedback on \No{}, 
    which was primarily recognized for its label stability.
    Many participants appreciated the limited movement of the labels, 
    mentioning that it made them ``\emph{easier to catch}'' 
    and that the ``\emph{length and direction did not change too much.}'' 
    However, a significant drawback noted by several participants
    was the excessive label occlusion,
    which made it difficult to read the labels when players were behind one another. 
    One participant stated that the labels ``\emph{overlapped a lot, sometimes [making it] impossible to recognize the numbers.}''

    \vspace{1mm}
    \item \Force{}.
    By contrast, 
    participants praised \Force{} for its ability to eliminate label occlusion but complained about the excessive jittering and abrupt movement of the labels.
    Several participants noted the absence of overlapping labels, with one stating ``\emph{almost no occlusion at all.}''
    However, some participants found 
    the unstable and unpredictable label movement distracting 
    and ``\emph{hard to follow,}''
    or even ``\emph{made me dizzy.}''
    Although we believe that improved results can be achieved through an enhanced force-based method, 
    it may still be prone to unpredictable movement, 
    as it does not take into account the future states of the labels.

    \vspace{1mm}
    \item \Ours{}.
    Participants appreciated \Ours{} primarily for its occlusion-free presentation and stable movement of labels. 
    Many participants praised the method's ability to avoid label overlap effectively, as one user stated, ``\emph{Overlap is avoided well.}''
    The stable movement was also well-received, as users found it easy to track labels, e.g., 
    ``\emph{movement of the labels is natural and easy to follow.}''     
    However, some participants noted drawbacks, such as labels occasionally moving out of the screen and the layouts are ``\emph{not so appealing visually.}''
    Future improvements can involve optimizing the visual aesthetics of the labels as one objective or considering users' gaze and attention to enhance the local layout of the labels. 
\end{itemize}

\subsection{Summary}
The user study demonstrated that resolving occlusions is crucial for reading labels in dynamic scenarios.
\Ours{}, the proposed method, was shown to effectively achieve this goal 
in both scenarios with fast-moving or many labels,
and assist users with various visual search tasks, such as identifying, comparing, and summarizing data on moving labels.
In addition, participants performed significantly better 
with \Ours{} than with \Force{} in comparing labels in the NBA dataset, suggesting that \Ours{} can help users read multiple fast-moving labels.
We attribute this to \Ours{}'s ability to stabilize label movements while resolving occlusions.
Participants' subjective feedback also suggested that \Ours{} can achieve both occlusion-free and stable movements of the labels.

\section{Discussion, Future Work, and Limitations}

\vspace{-1mm}
\para{AR Visualizations as Robots Without A Physical Body.}
Our RL method considers AR visualizations as robots without a physical body, which offers a key insight: if we can control robots to react and adapt to real-world environments, then we can do the same, or more, for AR visualizations. Although this concept is still in its early stages, it presents exciting opportunities for merging research in robotics with data visualization. By combining these two fields, we can create dynamic, interactive AR visualizations that adapt to real-world environments and assist users in making in-situ decisions. 
As we embark on this journey, we are excited to explore the potential of this innovative approach.

\para{Involving Human Feedback for RL-based Visualizations.}
The very purpose of visualizations is to aid humans in gaining insights into the data presented. 
Therefore, when developing RL methods for visualizations, 
it is essential to incorporate human feedback in the training process. By doing so, we ensure that the AR visualization not only fulfills the optimization objectives but also meets the user's requirements and expectations that are often difficult to quantify. 
Recent studies in large language models have highlighted the importance of human feedback in RL~\cite{ouyang2022training}.
Although we did not incorporate human feedback in the training process of our RL model as the first step, 
we are enthusiastic about exploring this direction in the future. 
Doing so would enable more advanced AR visualizations by, for example, incorporating aesthetics and level of detail in presenting the information.

\para{Generalizing to More Complex and Diverse Scenarios.}
Our experiments on the \STU{} datasets have already demonstrated the generalizability of our method to scenarios where the number of objects in the scene is continuously changing. 
Our method's data-driven nature enables us to apply it to other dynamic scenarios where objects or labels have different sizes, shapes, \cmo{or anchor points as long as we incorporate those factors into the observable states and train the network with relevant data}.
In the future, we aim to expand our method to more complex and varied scenarios, such as city spaces with pedestrians and cars, similar to controlling mobile robots~\cite{DBLP:conf/icra/ChenLKA19}.
\cmo{Furthermore, we are interested in exploring diverse action spaces that involve modifying the size, position (in the xyz coordinates), and opacity of the labels.}
Yet, this requires collecting additional movement datasets from the real world, which is beyond the scope of our current work. 
Nonetheless, our research lays the foundation for developing AR visualizations that can adapt to dynamic environments.

\para{Extending to Real AR scenarios.}
In this study, we utilized VR to simulate AR environments, 
allowing the view management system to acquire accurate object states, 
such as positions and velocities. 
In real AR environments, such states can be obtained through the use of sensors like LiDAR~\cite{lidar}.
However, these advanced sensors can be costly. 
A more cost-effective alternative is to use vision-based signals such as videos as input. For instance, Zhu et al.~\cite{DBLP:conf/icra/ZhuMKLGFF17} developed an RL-based approach for navigating robots to locate a specified target using solely visual inputs.
However, challenges such as real-time computing must be addressed in future research.

\para{Study Limitations.}
While the model experiment and user study 
have demonstrated that \system{} advances baselines,
several limitations exist in our study.
Firstly, we simplified the tasks by using cube shapes to represent humans and fixing the label size and viewer position. 
However, we believe that this is a necessary first step toward developing a more complex RL-based method, 
which is known to be more challenging to train compared to other machine learning models~\cite{lillicrap2015continuous}.
Secondly, we only evaluated users' performance on three fundamental visual tasks, while in-situ analytics with AR visualization can present more complex challenges. 
Finally, even though we followed exciting works~\cite{DBLP:journals/tvcg/LinYBP23, DBLP:journals/corr/abs-2209-00202, DBLP:journals/tvcg/YeCCWFSZW21, DBLP:journals/tvcg/ChuXYLXYCZW22},
we acknowledge that simulating AR in VR may not fully represent real-world scenarios. 
These limitations suggest potential areas for future research and improvements to \system{}.

\section{Conclusion}

We introduced \system{}, a novel RL-based method to address the challenges of managing AR label placements for moving objects.
To the best of our knowledge, this is the first study to utilize RL for managing AR labels of moving objects.
\system{} takes into account both current and future predicted environment states to make optimal decisions regarding label placements.
Our experimental results on two real-world trajectory datasets demonstrated that \system{} effectively learned the decision-making process and outperformed two baselines in reducing occlusions, line intersections, and movement distance of the labels. 
The user study further showed that \system{} improved user performance in various visual search tasks and achieved both occlusion-free and stable movements of the labels.
Overall, our work establishes a strong foundation for the development of more advanced and effective RL-based methods for AR visualizations.

\acknowledgments{
This research is
supported in part by 
NSF award III-2107328, NSF award IIS-1901030,
NIH award R01HD104969,
and the Harvard Physical Sciences
and Engineering Accelerator Award.
}

\bibliographystyle{abbrv-doi-hyperref}

\bibliography{template}
\end{document}